\documentclass[preprint,12pt,authoryear]{elsarticle}

\usepackage{amssymb}
\usepackage{ulem}
\usepackage{color}
\usepackage{amsmath}
\usepackage{graphicx}

\journal{International Journal of Solids and Structures}

\begin{document}

\begin{frontmatter}

\title{Band Structure Engineering of Coupled-Resonator Phononic Polyacetylene and Polyaminoborane} 

\author[ICF]{B. Manjarrez-Montañez} 
\affiliation[ICF]{organization={Instituto de Ciencias Físicas, Universidad Nacional Autónoma de México},
            city={Cuernavaca},
            postcode={62210}, 
            state={Morelos},
            country3
H. M. Simpson and P. J. Wolfe, ‘‘Young’s modulus and internal damping
in a vibrating rod,’’ Am. J. Phys. 43, 506–508 共1975兲.
={México}}
            
\author[ICF]{R. A. Méndez-Sánchez} 
\ead{mendez@icf.unam.mx}

\author[IFUNAM]{Y. Betancur-Ocampo} 
\affiliation[IFUNAM]{organization={Instituto de Física, Universidad Nacional Autónoma de México},
            city={Ciudad de México},
            postcode={04510}, 
            state={Cuidad de México},
            country={México}}

\author[IFBUAP]{A. Martínez-Argüello} 
\affiliation[IFBUAP]{organization={Instituto de Física, Benemérita Universidad Autónoma de Puebla},
            city={Puebla},
            postcode={72570}, 
            state={Puebla},
            country={México}}

\begin{abstract}

A methodology for constructing a quasi-one-dimensional coupled-resonator phononic metamaterial is presented. This is achieved through the design of artificial phononic analogs of two molecular structures: trans-polyacetylene and trans-polyaminoborane. The band structure of trans-polyacetylene is analyzed in relation to the Su-Schrieffer-Heeger (SSH) model, while that of trans-polyaminoborane is examined using the $\kappa$-deformed Dirac equation, both within a tight-binding framework. 
Additionally, the obtained finite realization of the artificial trans-polyacetylene exhibits topologically protected states.
\end{abstract}

\begin{graphicalabstract}
\includegraphics[width=\linewidth]{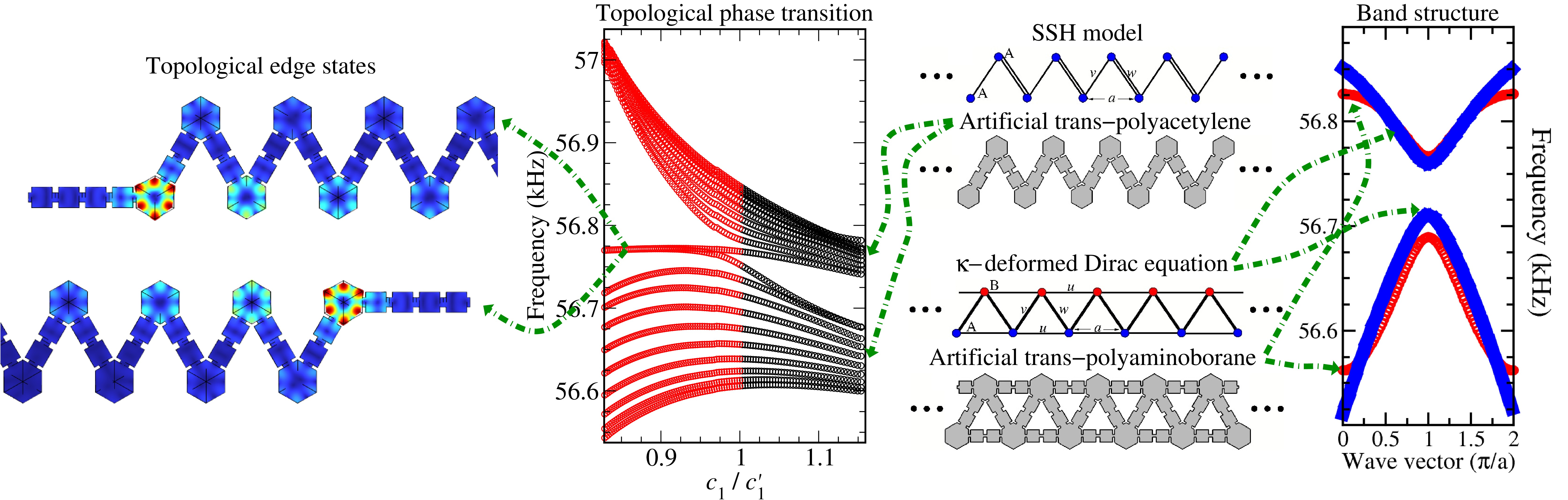}
\end{graphicalabstract}

\begin{highlights}
\item The methodology for constructing coupled-resonator phononic waveguides is presented. 
\item The SSH model is emulated with these waveguides with an error less than  1\%. 
\item Topologically protected states are observed in finite trans-polyacetylene chains. 
\item Artificial trans-polyaminoborane with first- and second-order hoppings is studied. 
\item The obtained band structure agrees with the results of the $\kappa$-deformed Dirac equation.
\end{highlights}


\end{frontmatter}



\section{Introduction}
\label{sec1}

Acoustic or phononic metamaterials are artificially engineered materials that exhibit mechanical properties not typically found in natural materials~\citep{Laude2015}. These unique properties arise not only from the constituent materials but also from their geometric design, where an engineered unit cell is repeated according to a specific pattern. As a result, elastic waves can be effectively controlled within these metamaterials~\citep{LuFengChen2009}.

The operational frequency range of phononic metamaterials spans from infrasound (Hz) to terahertz (THz) depending on the unit cell size. A common approach for designing metamaterials that operate within the frequency range from 0-100 kHz involves the machining of holes or notches in otherwise uniform beams or plates~\citep{JASA2002,JASA2005,Arreola-LucasEtAl2019,Lopez-ToledoBaezMendez-Sanchez,Martinez-GarciaEtAl2023}. In such cases, the unit cell consists of aluminum and air, with the latter acting as an effective vacuum. The properties of these metamaterials are determined by three key factors: (i) the filling fraction between air and aluminum, (ii) the geometric arrangement of holes or notches within the unit cell, and (iii) the spatial configuration of adjacent unit cells.
Despite significant advancements in the field, no universal methodology exists for designing and constructing metamaterials.

Coupled-resonator phononic metamaterials (CRPnMs) \citep{Ramirez-RamirezEtAl2020,Lopez-ToledoBaezMendez-Sanchez} are inspired by coupled-resonator optical waveguides \citep{Yariv1999}. In CRPnMs, resonators are coupled through finite phononic crystals \citep{KushwahaEtAl1993,JASA2002,JASA2005,KafesakiSigalasEconomou1995,SigalasEtAl,Manzanarez-MartinezRamos-Mendieta}. The construction of a CRPnM requires the following condition to be met:
\begin{itemize}
 \item[{\bf A}] {{When the normal-mode frequency of a resonator falls within the bandgap of the phononic crystal, the resonator modes couple through evanescent waves.}} \label{cond}
\end{itemize}
The importance of condition {\bf A} lies in its role in localizing the normal modes of the structure. 
When condition {\bf A} is satisfied, these modes become confined within the resonators and decay evanescently through the couplers, mimicking the behavior of atomic orbitals whose amplitudes diminish with distance. 
In this case, the overlap between two neighboring orbitals decreases with the number of unit cells in the coupler that separates them~\citep{Ramirez-RamirezEtAl2020}; this mirrors the decay of $\pi$ orbital overlap with increasing atomic separation. 
Such evanescent wave overlap is also characteristic of quantum-mechanical potential barriers, where similar decaying modes appear.  
As a result, many phenomena predicted in solid-state and molecular physics—often difficult to observe directly in those domains—can be emulated and studied using CRPnMs. 
Examples include the replication of spectral features such as Dirac cones and the realization of topological phase transitions, among others. 
Recently, we have developed a strategy for designing CRPnMs, which involves analyzing the components individually to meet condition {\bf A} before examining the entire system. 
This process is facilitated through comprehensive finite element simulations. 
Initially, it is essential to design a coupler with an appropriate bandgap, which is determined by calculating the band structure of the coupler's unit cell using Floquet (or Bloch) boundary conditions. 
A complete bandgap is a key point to prevent coupling between bending, compressional, and torsional waves. 
Often, it is necessary to analyze a finite phononic crystal, as the edge states of the coupler may intrude upon the desired bandgap. 
Next, the frequency spectrum of the resonator is assessed based on its geometric parameters, allowing for the selection of a parameter set that achieves the desired natural frequency within the coupler's bandgap. 
Subsequently, a supercell of the CRPnM, comprising both the resonator and couplers, is constructed, and its band structure is evaluated according to these parameters, with the expectation that a new band will emerge within the chosen bandgap. 
Fine adjustments to the parameters may also be required to achieve the bandgap with the intended characteristics. 
Finally, for experimental implementation, a finite version of the CRPnM must be designed and analyzed, incorporating terminators --couplers with multiple unit cells-- to mitigate border effects. 
Additionally, it is important to consider the spacing between the resulting resonances, as there is a minimum resolvable separation. Recent experiments suggest that a separation of approximately 40~Hz is required to effectively distinguish closely spaced resonances~\citep{Martínez-ArguelloEtAl2022}.

Polyacetylene is a polymeric molecule notable for its significant electronic properties~\citep{ShirakawaEtAl1977,ChiangEtAl1977}. As a highly conductive polymer, it serves as a fundamental component in the realm of organic semiconductors~\citep{GuoFacchetti2020}. The structure of polyacetylene features a chain of carbon atoms linked by alternating single and double bonds, with each carbon atom bonded to a hydrogen atom. The Su-Schrieffer-Heeger (SSH) model provides a theoretical framework for analyzing this polymer, revealing that polyacetylene exhibits topologically protected states~\citep{SuSchriefferHeeger}. This characteristic, among others, has sparked interest in the emulation of polyacetylene in various studies~\citep{CoutantEtAl,FanEtAl,LiEtAl,BarriosVargasEtAl,PadlewskiEtAl,Meier2016,MandalKar,DALPOGGETTO,GangulyMaiti}. Polyacetylene exists in two isomeric forms: cis- and trans-polyacetylene, which correspond to distinct borders of graphene. 
We have shown recently that the cis isomer, particularly when considering first- and third-nearest neighbors, can be effectively emulated using CRPnM~\citep{Betancur-OcampoManjarrez-MontanezMartinez-ArguelloMendez-Sanchez}. Notably, it has been discovered that this configuration satisfies a modified Dirac equation, leading to the emergence of two distinct topological phases.

The trans-polyaminoborane molecule is similar to trans-polyacetylene, where alternating boron and nitrogen atoms substitute the carbon ones. 
This structure, which incorporates first- and second-nearest-neighbors hoppings, obeys a $\kappa$-deformed Dirac equation~\citep{MajariEtAl2021}. 
While a photonic realization was proposed there, achieving higher couplings in that domain becomes exceedingly challenging, if not nearly impossible~\citep{StegmanEtAl2017}. As it will be demonstrated in the following sections, the engineering of first- and second-nearest neighbors in the coupled-resonator phononic variant of these molecules is a relatively straightforward process.

This paper presents an application of the methodology outlined in the previous paragraphs for designing coupled-resonator phononic metamaterials. The focus will be on two specific molecules: coupled-resonator phononic trans-polyacet\-y\-lene (CRPnTPA) and coupled-resonator phononic trans-poly\-aminoborane (CRPnTPB), examining both first- and second-nearest neighbor interactions. The subsequent section will detail the design of the resonator and coupler, along with the frequency spectrum of the former and the band structure of the latter. Section~\ref{Sec.SSH1} will analyze the CRPnTPA with first-nearest neighbors, comparing the findings to the SSH model, showing a very good agreement. In Section~\ref{Sec.kappaDeformed}, the study will extend to the CRPnTPB, incorporating both first- and second-nearest neighbors, with numerical results compared to the phononic tight-binding model (TB model) for the $\kappa$-deformed Dirac equation, revealing excellent agreement. The finite versions of CRPnTPA and CRPnTPB will be examined in Section~\ref{Sec.finite}, showing both the spectra and wave amplitudes. The spectrum of the CRPnTPA, as a function of the geometric parameters, indicates the existence of a topological phase transition.  
The paper will conclude with a summary of the findings and two appendices discussing the phononic TB model and the $\kappa$-deformed Dirac equation.

\section{Design of coupled-resonator phononic metamaterials}

\begin{figure}[tb]
    \centering
    \includegraphics[width=0.9\columnwidth]{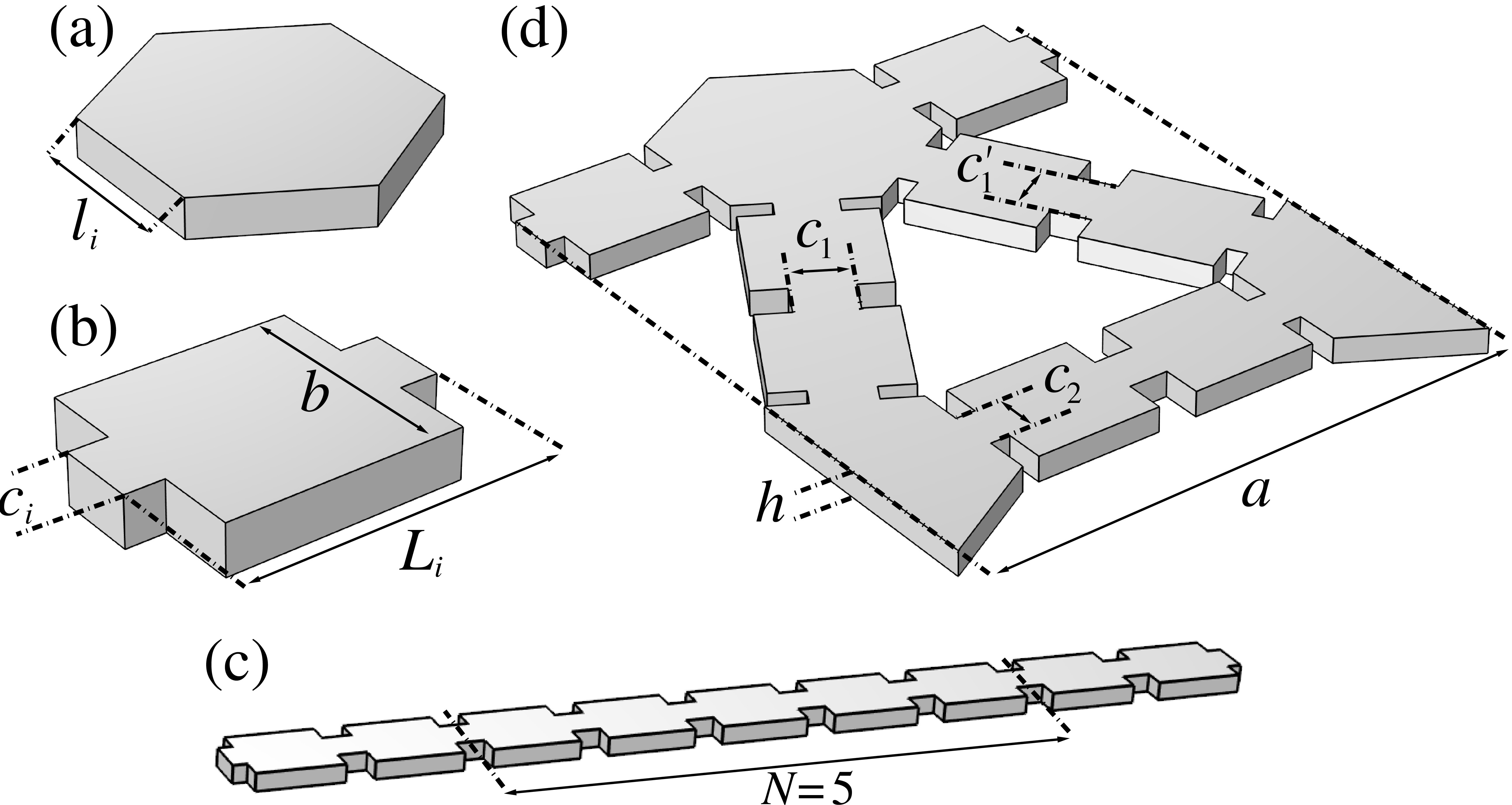}
    \caption{Coupled-resonator phononic metamaterial parts: (a) hexagonal resonator  of side $l_i$ and thickness $h$, (b) unit cell of the coupler composed of two cuboids of $(L_i-b)\times c_i\times h$ and a square plate of $b\times b \times h$,  (c) one-dimensional phononic crystal with $N=5$ unit cells, (d) unit cell of the coupled-resonator phononic polyacetylene. The geometrical parameters are $l_1= 34.1$~mm, $l_2= 34.2$~mm, $L_1= 41$~mm, and the combinations of  $c_1$ and ${c_1}^\prime$ = $1, 15, 16$~mm, $L_2 = 41.035$~mm, $b = 31$~mm, $c_2 = 0.5, 5, 10, 12$~mm and $h=6.35$~mm.
    }
    \label{Fig.UnitCell}
\end{figure}

In this section, the spectral properties of the unit cell composing the CRPnM are presented. The CRPnM is machined from a single aluminum plate. Its unit cell (see Fig.\ref{Fig.UnitCell} (d)) consists of hexagonal plates with side lengths $l_i$ ($i=1,2$) and thickness $h$ (see Fig.\ref{Fig.UnitCell} (a)). Additionally, it includes a two-cell finite phononic crystal, where each unit cell is composed of two cuboids with dimensions $c_i\times (L_i-b) \times h$ and a square plate with side length $b$ and thickness $h$ (see Fig.\ref{Fig.UnitCell} (b)). A finite phononic crystal comprising five unit cells is highlighted in the structure shown in Fig.\ref{Fig.UnitCell} (c).

\begin{figure}[htb!]
    \centering
    \includegraphics[width=\linewidth]{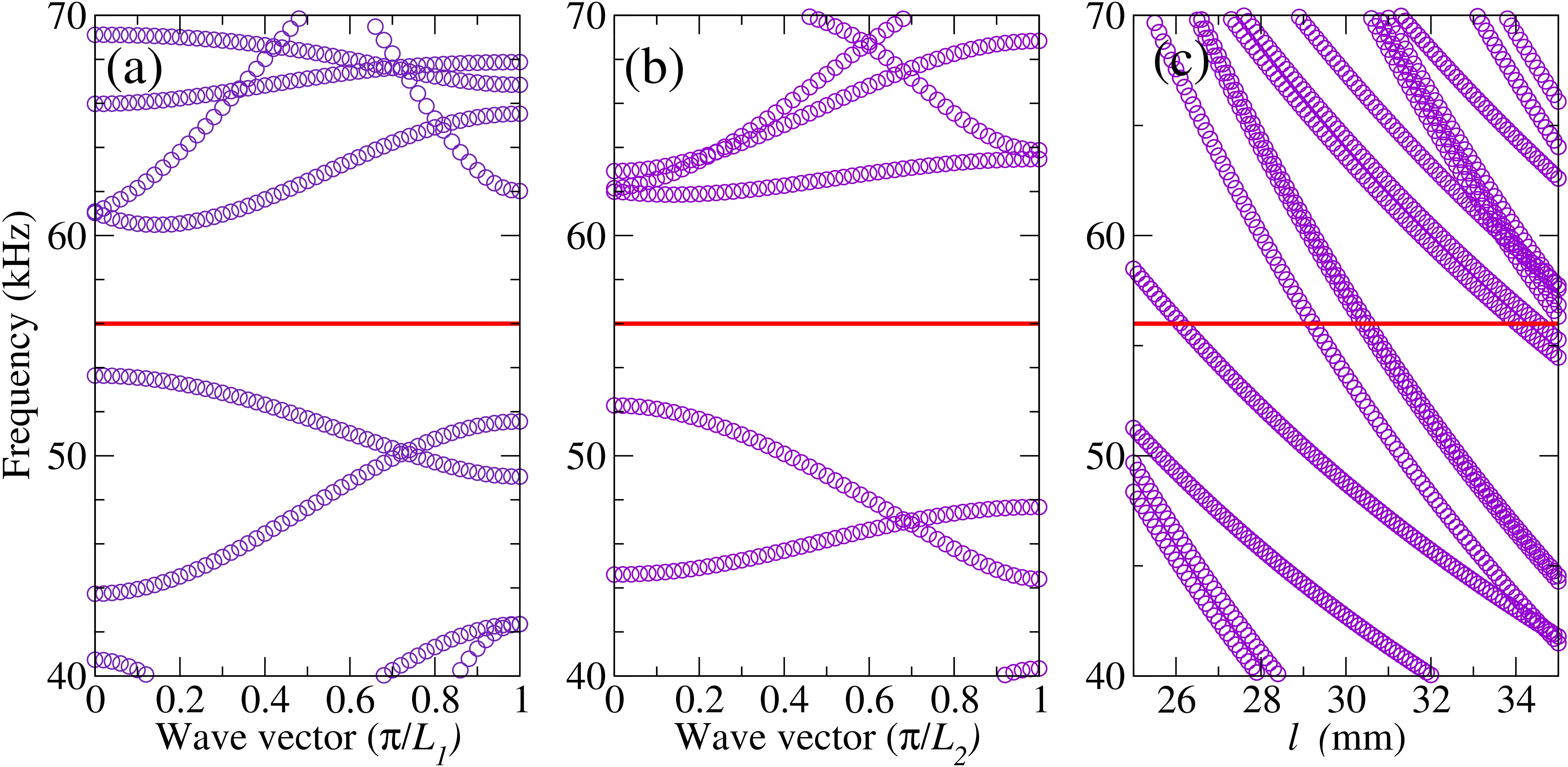}
    \caption{(a) Band structure of the phononic crystal with $L_1=41$~mm and (b) with $L_2=41.035$~mm (c) Normal-mode frequencies of the hexagonal plate as a function of the hexagon side. For $l\approx 34$~mm the resonator eigenfrequency at approximately 56 kHz is located within the phononic bandgaps given in (a) and (b).}
    \label{Fig.PnCBandStructure}
\end{figure}

\begin{figure}[htb!]
    \centering
    \includegraphics[width=\linewidth]{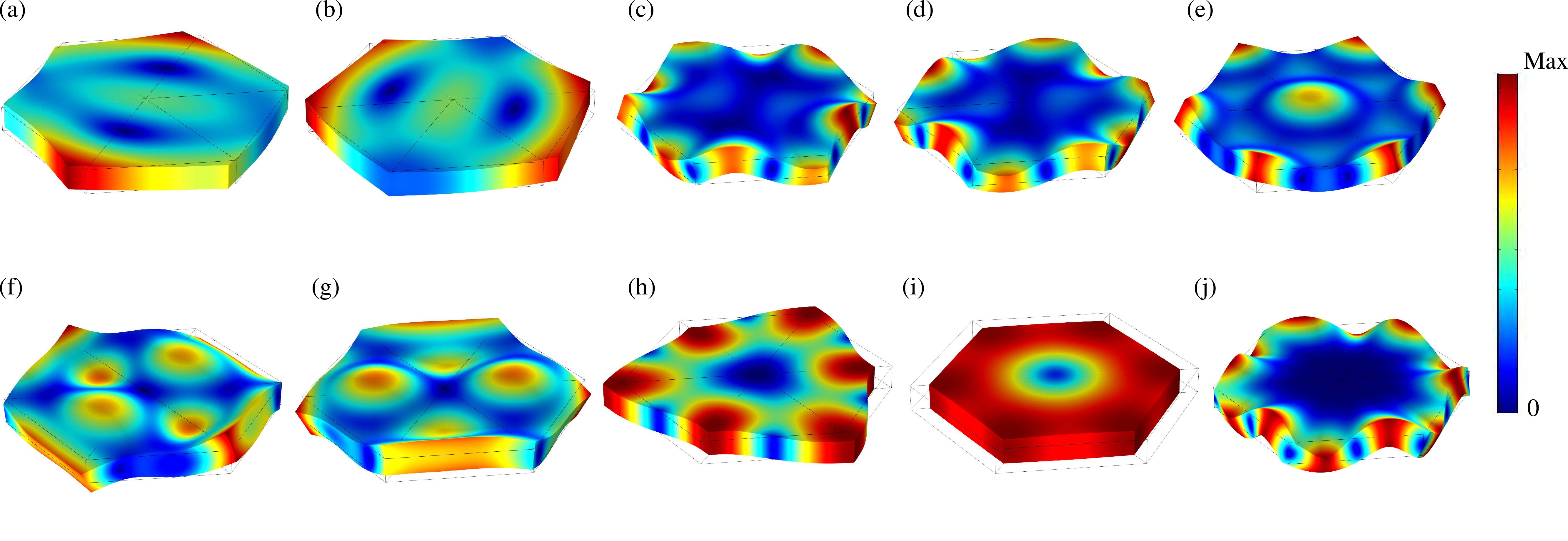}
    \caption{Some normal-mode wave amplitudes of the resonator given in Fig.~\ref{Fig.UnitCell} (a) with $l=34.1$~mm. The frequencies are (a) and (b) 42 893 Hz, (c) and (d) 43 364 Hz, (e) 46 285 Hz, (f) and (g) 46 550Hz, (h) 55 898 Hz, (i) 56 716 Hz, and (j) 58 768 Hz.}
    \label{Modes}
\end{figure}

To determine the band structure of a phononic crystal, one can apply Bloch's theorem, allowing the analysis of a single unit cell instead of the entire crystal. The numerical calculations are performed using the commercial software COMSOL Multiphysics.
Figure~\ref{Fig.PnCBandStructure} (a) displays the band structure of the phononic crystal whose unit cell is depicted in Fig.~\ref{Fig.UnitCell} (b). 
These phononic bands are computed using the finite element method (FEM) within the unit cell, applying free boundary conditions on all surfaces except at the left and right rectangular boundaries. At these surfaces, Floquet boundary conditions are imposed, meaning that the displacement components $\vec u_\mathrm{destiny}$ on the left surface are linked to those on the right surface, $\vec u_\mathrm{source}$, through
\begin{equation}
   \vec u_\mathrm{destiny}=e^{i k L_i} \vec u_\mathrm{source},
   \label{Bloch}
\end{equation}
where the wavenumber $k$ is used as a parameter to determine the allowed bands that satisfy the boundary conditions of Eq.~(\ref{Bloch}). In Fig.~\ref{Fig.PnCBandStructure}(b), a different band structure is presented using an alternative set of parameters for the coupler unit cell. The focus then shifts to the resonator. In Fig.~\ref{Fig.PnCBandStructure}(c), the normal-mode frequencies of a hexagonal plate with side length $l$ are determined under free boundary conditions. For $l\approx 34$~mm, a normal-mode frequency of approximately 56~kHz is obtained. This frequency falls within the phononic crystal bandgaps depicted in Figs.~\ref{Fig.PnCBandStructure}(a) and (b). The corresponding wave amplitude, an in-plane mode at this frequency, is illustrated in Fig. \ref{Modes}(h). Other modes shown in the same figure can be utilized to construct the CRPnM provided their respective normal-mode frequencies align with the coupler bandgap. For instance, mode (i) is suitable for a resonator size of 34.5~mm.  
The integration of a resonator with the phononic crystals results in the localization of the resonator modes, which vanish evanescently throughout the phononic crystal.

\section{Coupled-resonator phononic trans-polyacetilene}
\label{Sec.SSH1}

\begin{figure}[ht!]
    \centering
    \includegraphics[width=\linewidth]{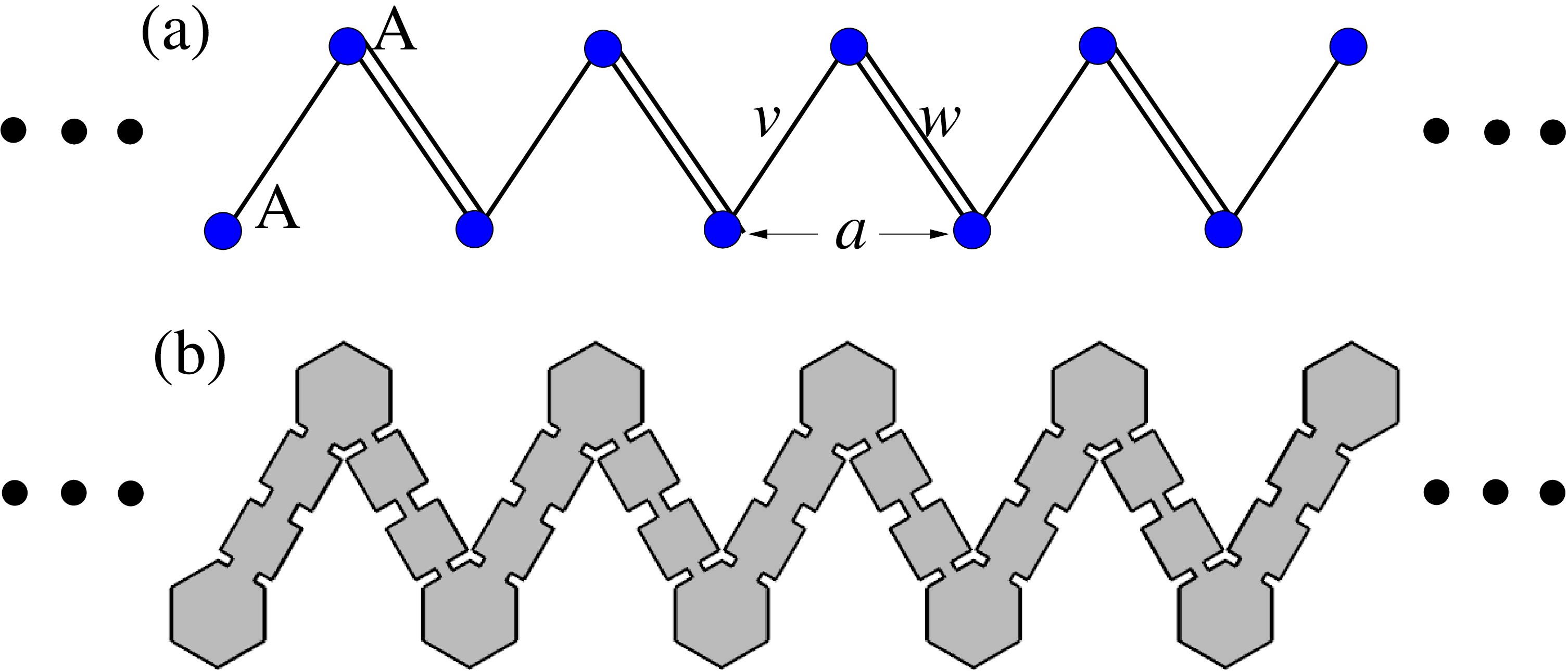}
    \caption{Trans-polyacetylene chains. (a) Pictorial representation of the Su–Schrieffer–Heeger model changing the first-nearest-hopping parameters $v$ and $w$. (b) Phononic trans-polyacetylene chain.}
    \label{Fig.SSH}
\end{figure}

In this section, we analyze the coupled-resonator phononic trans-poly\-acet\-y\-lene (CRPnTPA). A schematic representation of trans-polyacetylene is provided in Fig.~\ref{Fig.SSH} (a). 
The trans-polyacetylene has two distinct nearest-neighbor hopping parameters, $v$ and $w$ so two different types of couplers are used in the CRPnTPA. 
Since only a single type of atom is present, all resonators will have the same size in the CRPnTPA. 
Trans-polyacetylene exhibits a transition between a trivial and a topological phase, governed by the ratio $v/w$~\cite{SuSchriefferHeeger}. 
When $w<v$, the system remains in a trivial phase. 
A linear dispersion relation emerges when $v=w$, for the critical value in the transition. 
In contrast, for $w>v$, the system enters a topological phase. 
Details on the tight-binding approach of the phononic molecule are given  in~\ref{App.TPA-B}.
The well-known topological phase transition observed in trans-polyacetylene can also be reproduced in CRPnTPA by varying the parameters $c_1$ and $c_1'$, while keeping all other parameters fixed.
The unit cell of the CRPnTPA is depicted in Fig.~\ref{Fig.UnitCell}(d), where the chosen values for $c_1$ and ${c_1}^\prime$ as $1$, $15$, $16$~mm, corresponding to the hopping parameters $v$ and $w$, respectively. To eliminate second-nearest-neighbor couplings, the width of the smaller cuboids was set to zero, i.e., $c_2=0$~mm.

\begin{figure}[ht!]
    \centering
    \includegraphics[width=\linewidth]{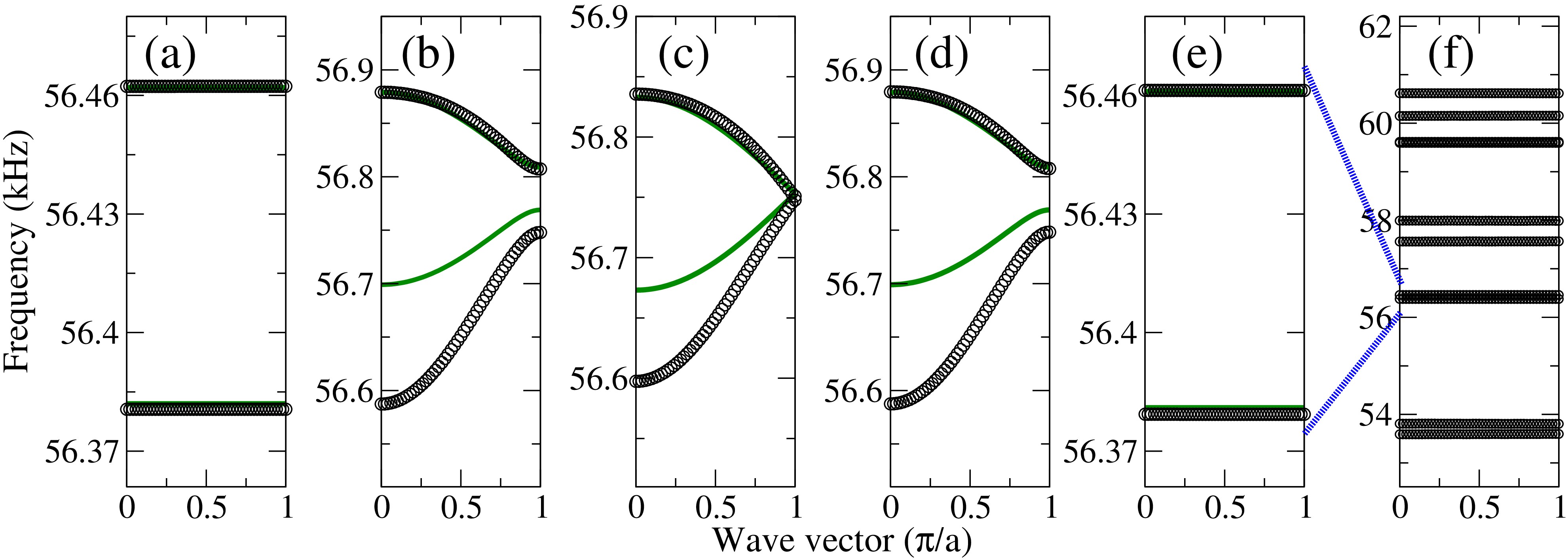}
    \caption{Band structure of the coupled-resonator phononic trans-polyacetylene. The black points correspond to finite-element calculations. The green curves are obtained from the SSH model (Eq.~(\ref{Eq.Frequencies})). The widths of the small cuboids are (a) $c_1'=1$~mm and $c_1=15$~mm ($w=0$). (b)  $c_1'=15$~mm and $c_1=16$~mm ($v>w$). (c) $c_1'=c_1=15$~mm ($v=w$). (d)  $c_1'=16$~mm and $c_1=15$~mm ($v<w$). (e) $c_1'=15$~mm and $c_1=1$~mm ($v=0$). (f) Wide zoom of our interest bands.}
    \label{shh}
\end{figure}

The band structure of the CRPnTPA is presented in Fig.~\ref{shh} (a)–(e) for various values of $c_1$ and ${c_1}^\prime$. 
The results reveal two bands separated by a gap when $c_1 \ne {c_1}^\prime$, which merge into a Dirac cone when $c_1={c_1}^\prime$. 
The topological phase is depicted in Figs.~\ref{Fig.SSH} (c) and (d), demonstrating a very good agreement with the SSH model. 
Although the numerical results for the lower band do not seem to agree perfectly, one should notice that a wide zoom in the frequency scale is performed, as seen in Fig.~\ref{shh} (f).   
Indeed the percentage error between the model predictions and numerical simulations remains below 0.2\%. 
The observed differences come from the quite small number of cells used in the finite phononic crystals that yield second nearest-neighbor hoppings and a non-orthogonal tight-binding basis~\citep{Santiago-GarciaMendez-SanchezSadurni}.

\section{Coupled-resonator phononic trans-polyaminoborane}
\label{Sec.kappaDeformed}

\begin{figure}[ht!]
    \centering
    \includegraphics[width=\linewidth]{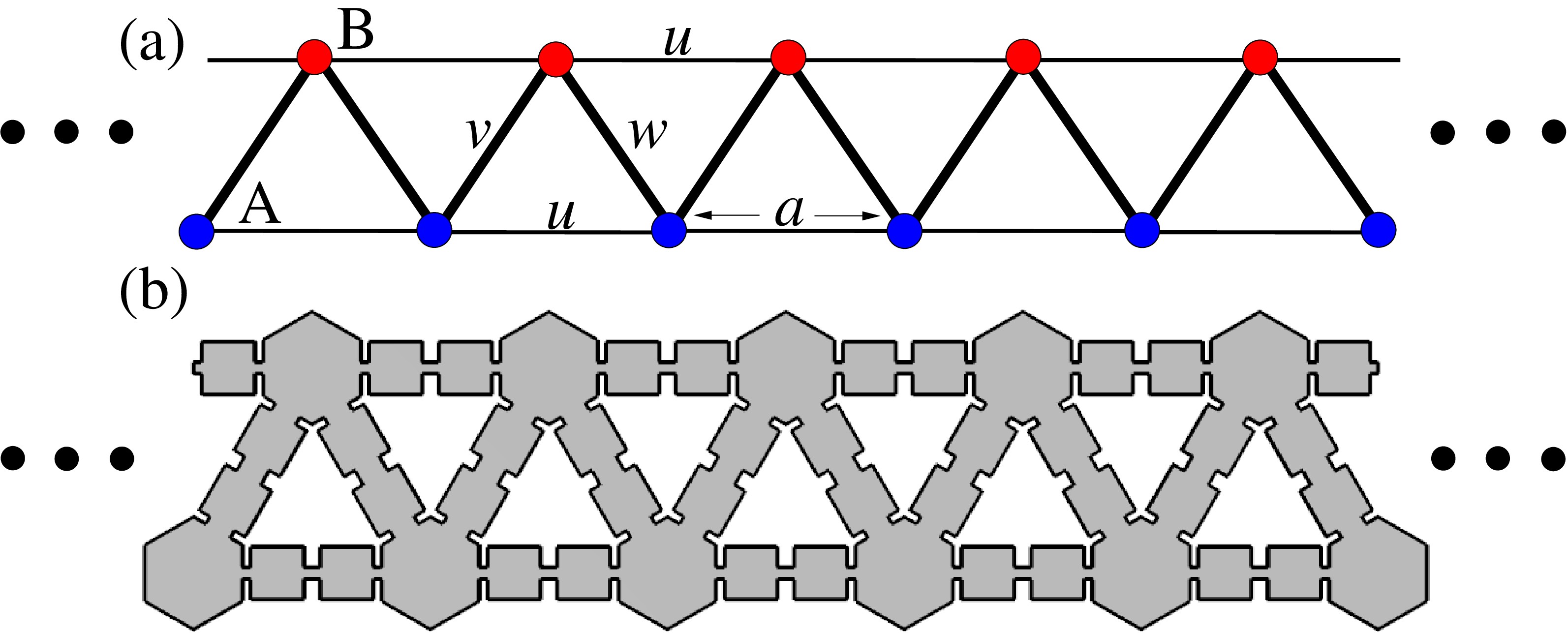}
    \caption{(a) Trans-polyamynoborane chain with first- and second-nearest neighbors. The hopping parameters $u=w$ and $u$ respectively. The energies sites A and B are differents. (b) Coupled-resonator phononic trans-polyaminoborane with first- and second-nearest neighbors..}
    \label{Fig.TPBMolecule}
\end{figure}

\begin{figure}[ht!]
    \centering
    \includegraphics[width=\linewidth]{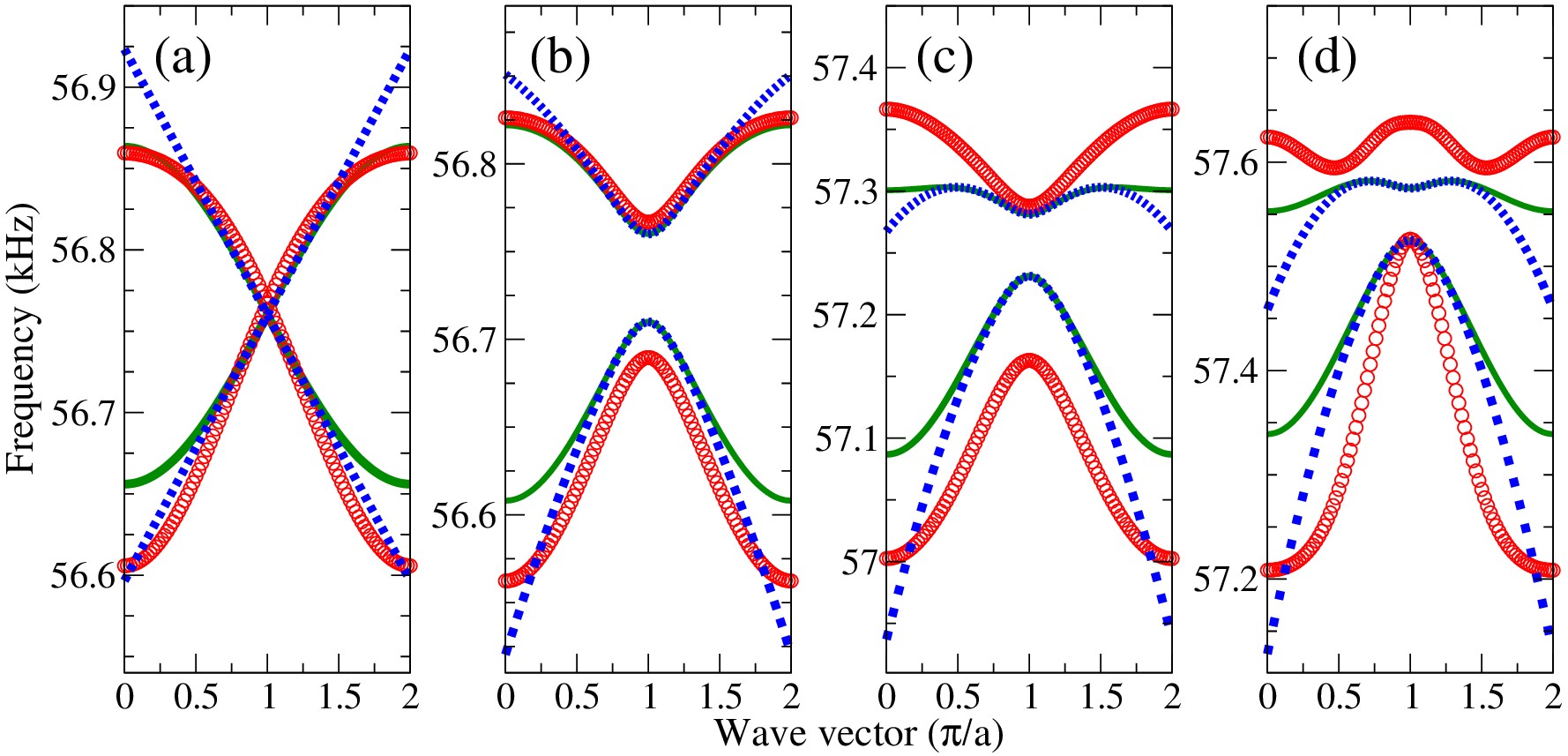}
    \caption{Band structure of the artificial trans-polyaminoborane. The red points correspond to finite-element calculations. The green curves are obtained from Eq.~(\ref{Eq.Frequencies}). The blue dotted line correspond to $\kappa$-deformed Dirac equation from Eq.~(\ref{eq:HKD}): (a) $f_A = f_B = 56760$~Hz, $v = w=52$~Hz, and $u = 0$~Hz. (b) $f_A = 56700$~Hz, $f_B = 56750$~Hz, $u =w= 52$~Hz, and $-u = 5$~Hz. (c) $f_A = 57130$~Hz, $f_B = 57180$~Hz, $u = w=52$~Hz, and $-u = 15.6$~Hz. (d) $f_A = 57473$~Hz, $f_B = 57523$~Hz, $v = w=52$~Hz, and $-u = 26$~Hz.}
    \label{Fig.TPBBands}
\end{figure}

In this section, the trans-polyaminoborane is emulated using coupled-resonators phononic systems.
A representation of this molecule is shown in Fig.~\ref{Fig.TPBMolecule}(a), while the design of the coupled-resonator phononic trans-poly\-aminoborane (CRPnTPB) appears illustrated in Fig.~\ref{Fig.TPBMolecule}(b).
Since atoms A and B are distinct, two different site frequencies are required.
This is achieved by employing two hexagonal resonators with sizes $l_1=34.1$~mm and $l_2=34.2$~mm.
The nearest-neighbor couplings are kept constant, whereas the next-nearest-neighbor couplings are varied.
In the physical structure, this is implemented fixing $c_1$ and varying $c_2$ (See Fig.~\ref{Fig.UnitCell}). 
Since the resonators have different sizes, the upper chain yields a different length than the lower chain of resonators in Fig.~\ref{Fig.TPBMolecule}(b). 
This is solved by adjusting the values of $L_i$ to get a unit cell size $a$ in both chains. 
At this stage changes in sizes are pretty small, so a redesign of the complete molecule is not necessary. 
Figure~\ref{Fig.TPBBands} displays both numerical and analytical results for the phononic band structures of the CRPnTPB, obtained via finite element simulations (red circles), the tight-binding approach (green lines), and the $\kappa$-deformed Dirac equation (blue dots). In subfigures (a)–(d), the width $c_2$ is systematically varied to tune the next-nearest-neighbor coupling within the chain, while the width $c_1$ is held constant.
In Fig.~\ref{Fig.TPBBands}(a), the site frequencies (or resonator sizes) are identical, resulting in a linear dispersion relation.
Regardless of the value of $c_2$, the CRPnTPB band structure exhibits a gap.
As $c_2$ increases, the upper band deforms and changes in width, while the lower band expands.
This behavior aligns with the tight-binding model and the approximation to the $\kappa$-deformed Dirac equation at the center of the cone~\cite{MajariEtAl2021}; this approximation is described in~\ref{App.Dirac}.
The phononic tight-binding scheme for the CRPnTPB is detailed in~\ref{App.TPA-B}. 
At very high deformations the CRPnTPB band structure shows deformations that are captured neither by the tight-binding model nor by the $\kappa$-deformed Dirac equation.

\section{Finite molecules and the topological invariant}
\label{Sec.finite}

\begin{figure}[ht!]
    \centering
    \includegraphics[width=\linewidth]{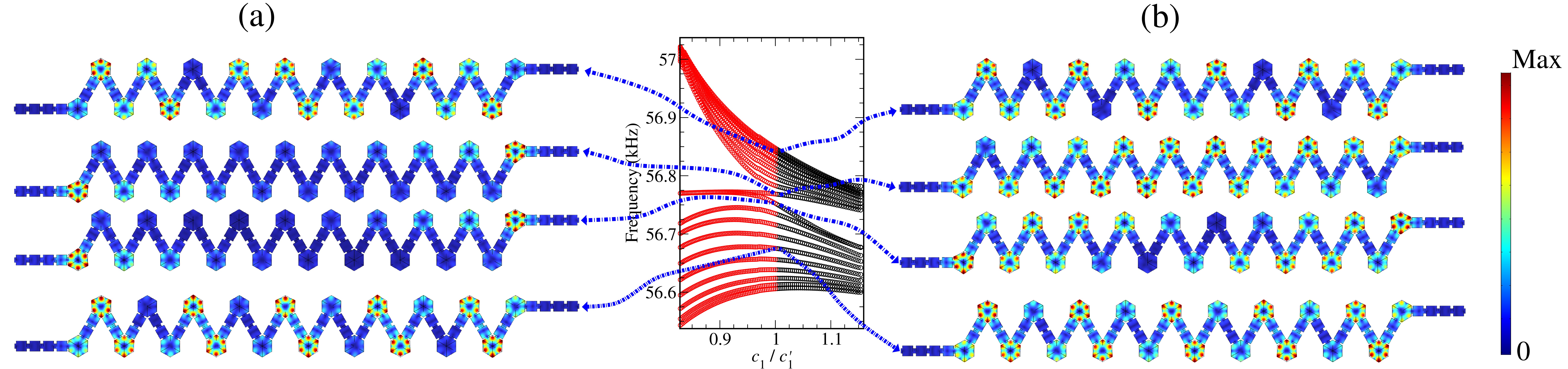}
    \caption{Frequency spectrum and normal wave amplitudes of a finite coupled-resonator phononic trans-polyacetylene  for intercell hopping amplitude $w=15$~mm as a function the intracell hopping amplitude $v$. Here $v < 15$~mm ($v>w$) corresponds to the topological (trivial) phases. (a) Topological states located in the middle gap for the ratio $c_1^\prime>c_1$. (b) Bulk states for the ratio $c_1^\prime<c_1$.}
    \label{Fig.Solitons}
\end{figure}

Now, the finite coupled-resonator phononic trans-polyacetylene (CRPnTPA) will be analyzed, as it exhibits topologically protected edge states.
In contrast, the finite coupled-resonator phononic trans-polyaminoborane does not present significant normal-mode wave amplitudes; therefore they will be omitted.
Figure~\ref{Fig.Solitons} displays selected normal-mode wave amplitudes of the finite CRPnTPA.
Case (a) at the left corresponds to the topological phase, where $c_1 < c_1'$. 
Case (b) at the right, gives states for the trivial phase ($c_1' < c_1$). 
At the center of the figure, the spectrum shows a topological phase transition: for  $c_1 < c_1'$ the number of edge states remains fixed for adiabatic changes, revealing a topological invariant~\citep{asboth2016short}.

\section{Conclusions}

In this study, we have introduced a methodology for constructing coupled-resonator phononic metamaterials incorporating first- and second-nearest-neighbor couplings.
This was achieved by connecting resonators through finite phononic crystals (couplers).
A coupled-resonator phononic metamaterial emerges when the normal-mode frequency of the resonator falls within the bandgap of the coupler.
We employed this approach to emulate key features of the electronic spectra of trans-polyacetylene and trans-polyaminoborane. 

The coupled-resonator phononic trans-polyacetylene exhibits a phase transition between trivial and topological phases, with a Dirac cone appearing at the critical point.
The topological phase is evidenced by the constant number of edge states within the bandgap, indicating the presence of a topological invariant. 
The obtained band structures show excellent agreement with those predicted by the tight-binding model for the Su-Schrieffer-Heeger Hamiltonian. 
The numerical normal-mode wave amplitudes for the coupled-resonator phononic trans-polyacetylene confirms that states in the trivial phase are extended, whereas the states inside the bandgap in the topological phase are localized at the boundaries of the chain. 
On the other hand, the coupled-resonator phononic trans-polyaminoborane displays a deformation of the upper band as a function of the second-nearest-neighbor coupling, consistent with the results of the tight-binding realization of the $\kappa$-deformed Dirac equation. 
At very high second-nearest-neighbor hopping strengths in the CRPnTPB, the band structures exhibit deformations that are not captured by the analytical models.

The methodology presented here demonstrates that the electronic spectrum of one-dimensional molecules with first- and second-nearest-neighbor hoppings can be effectively emulated using coupled-resonator phononic molecules.
Furthermore, this approach opens a new avenue in phononic metamaterial research, as it enables the emulation of various systems governed by the tight-binding model using coupled-resonator phononic metamaterials with high control over high-order neighbor hoppings.

\section*{Acknowledgments}
RAMS and YBO were supported by PAPIIT-UNAM under projects IN108825, IA106223, and IA102125. 
This work was supported by SECIHTI under Grant No. CF-2023-G-763.
BMM is grateful to SECIHTI for financial support through a Ph. D. fellowship. AMMA acknowledges financial support from SECIHTI under the program ``Estancias Posdoctorales por M\'exico 2022".

\appendix
\section{Phononic tight-binding model for the Su-Schrieffer-Heeger Model and trans-polyaminoborane}
\label{App.TPA-B}

To write the equations of the phononic tight-binding model, let us consider a mix of the chains given in the upper parts of Figs.~\ref{Fig.SSH} and~\ref{Fig.TPBMolecule}.  
In this case, two site frequencies $f_A$ and $f_B$ are taken.  
Also, two parameters for first-nearest-neighbors hoppings ($u$, $w$) and one for the second-nearest-neighbors hoppings ($u$) are used. 
The Hamiltonian for the unit cell of length $a$ is given by
\begin{equation}
H(ka)=
    \begin{pmatrix}
        f_A+2u\textrm{cos}(ka) & v+w \mathrm{e}^{-ika}\\
        v+w \mathrm{e}^{ika} & f_B+2u \textrm{cos}(ka)
    \end{pmatrix}. 
    \label{model}
\end{equation}
The solutions of the eigenvalue equation are
\begin{equation}
    f_\pm(ka)=\frac{1}{2} \Big( f_c + 4u \cos ka \pm e^{-ika}\sqrt{e^{2ika}(4 v^2+4w^2+{f_c}^2 + 8vw \cos ka) } \Big)
\label{Eq.Frequencies}
\end{equation}
where $f_c=(f_A+f_B)$. 
Eq.~(\ref{Eq.Frequencies}) gives a band structure that predicts that of the coupled-resonator phononic trans-polyaminoborane. 
The results for the trans-polyacetylene are obtained taking $f_A=f_B$, and $u=0$.
In table~\ref{Tabla} the parameters needed to obtain the results for the Su-Schrieffer-Hegger model and the trans-polyaminoborane are given. 

\begin{center}
\begin{table}
{\caption{\label{Tabla} Parameter values for the tight-binding models under consideration.}}
\begin{tabular}{|c|c|}
\hline
  Su-Schrieffer-Heeger  & Trans-polyamynoborane \\
  Model & ($\kappa$-deformed Dirac equation) \\
\hline
   $v$ $\neq$ $w$  & $v$ = $w$ \\
\hline   
   $f_A$ = $f_B$ & $f_A$ $\neq$ $f_B$ \\
\hline
   $u$ = $0$ & $u$ $\geq$ $0$ \\
\hline
\end{tabular}
\end{table}
\end{center}

\section{The $\kappa$-deformed Dirac equation}
\label{App.Dirac}
The tight-binding approach has been successfully applied to emulate, for instance, relativistic wave equations by means of ad hoc engineering of the corresponding dispersion relations. \cite{MajariEtAl2021} find a $\kappa$-deformed Dirac equation, in units of $\hbar = c = 1$ with $\hbar$ and $c$, respectively, the Planck constant and the speed of light, given by
\begin{equation}
\Big[ \mathrm{i} \gamma^{0} \Big( \partial_{0} + \frac{\mathrm{i} a}{2} \nabla^{2} \Big) + \mathrm{i} \gamma^{i} \partial_{i} + m \Big] \psi = 0 ,
\label{eq:Kdeformed}
\end{equation}
where $\gamma^{\mu}$ are the usual gamma matrices, $p_{i} = \mathrm{i} \partial_{i}$ is the particle momentum with mass $m$, $a$ is the deformation parameter, and $\psi$ is the wavefunction. From equation~(\ref{eq:Kdeformed}), the corresponding Hamiltonian is
\begin{equation}
H = \alpha \cdot p + \frac{a}{2} \nabla^{2} + m \beta ,
\label{eq:HKdeformed}
\end{equation}
where in the representation of Dirac matrices in $1 + 1$ dimensions are $\alpha_{1} = \sigma_{1}$ and $\beta = \sigma_{3}$, with $\sigma_{1,3}$ the corresponding Pauli matrix.
Notice that when $v = w$, $\varepsilon_{1} = -\varepsilon$, and $\varepsilon_{2} = \varepsilon$, the tight-binding Hamiltonian~(\ref{model}) reduces to
\begin{equation}
H(ka) =
\left(
\begin{array}{cc}
-\varepsilon + 2\, u \cos ka & w + w \mathrm{e}^{ -\mathrm{i} ka } \\ 
w + w \mathrm{e}^{ \mathrm{i} ka } & \varepsilon + 2\, u \cos ka
\end{array} \right) ,
\label{eq:HKD}
\end{equation}
whose dispersion relation is
\begin{equation}
E(ka) = 2 u \cos ka \pm \sqrt{2 w^{2} (1 + \cos ka) + \varepsilon^{2} } .
\end{equation}
A first-order expansion around $ka = \pi$ of Hamiltonian~(\ref{eq:HKD}) is given by
\begin{equation}
H_{\mathrm{approx}}(ka) \approx
\left(
\begin{array}{cc}
-\varepsilon + u\,(ka - \pi)^{2} - 2\, u & w\, (ka - \pi) \\ 
w\, (ka - \pi) & \varepsilon + u\,(ka - \pi)^{2} - 2\, u
\end{array} \right) .
\label{eq:HKDapprox}
\end{equation}
Notice that by setting $ka + \pi \rightarrow p_{1}$, Hamiltonian~(\ref{eq:HKDapprox}) can be written as
\begin{equation}
H_{\mathrm{approx}} = u p_{1}^{2}\, \textbf{1}_{2} + w\, p_{1} \sigma_{1} - \varepsilon \sigma_{3} - 2u\, \textbf{1}_{2} = H_{0} + V ,
\label{eq:Happrox}
\end{equation}
where $\textbf{1}_{2}$ is the unit matrix of dimension 2 and $V = - 2u\, \textbf{1}_{2}$. Finally, notice that Hamiltonian~(\ref{eq:HKdeformed}) can be mapped to $H_{0}$ of equation~(\ref{eq:Happrox}) after the formal change $-u = \frac{a}{2}$, $w = 1$, and $\varepsilon = m$. For comparison purposes, in Figure~\ref{Fig.TPBBands} the bands obtained from~(\ref{eq:HKDapprox}) are also included. That is, the coupled-resonator phononic trans-polyaminoborane agrees with Hamiltonian~(\ref{eq:HKDapprox}), which in turn obeys the $\kappa$-deformed Dirac equation~(\ref{eq:HKdeformed}). 
To our knowledge, this result has not been achieved in photonic realizations.

\bibliographystyle{elsarticle-harv} 

\end{document}